\documentclass[a4paper]{article}
\usepackage{amscd,amsmath,amssymb}

\newcommand{\mmu}{\mathbf{u}}
\newcommand{\mmbu}{\mathbf{\bar u}}
\newcommand{\mmU}{\mathbf{U}}
\newcommand{\mmbU}{\mathbf{\overline U}}
\newcommand{\mf}{\mathbf{f}}
\newcommand{\mpsi}{\boldsymbol{\psi}}
\newcommand{\mbpsi}{\bar{\boldsymbol{\psi}}}
\newcommand{\mlambda}{\boldsymbol{\lambda}}
\newcommand{\mblambda}{\boldsymbol{\bar\lambda}}
\newcommand{\mLambda}{\boldsymbol{\Lambda}}
\newcommand{\mbLambda}{\overline{\boldsymbol{\Lambda}}}

\newcommand{\p}[1]{(\ref{#1})}

\newcommand{\cD}{{\cal D}}

\newcommand{\bT}{{\overline T}{}}
\newcommand{\bG}{{\overline G}{}}
\newcommand{\bZ}{{\overline Z}{}}

\newcommand{\bD}{{\overline D}{}}

\newcommand{\bQ}{{\overline Q}{}}
\newcommand{\bS}{{\overline S}{}}

\newcommand{\bLambda}{{\overline \Lambda}{}}
\newcommand{\bOmega}{{\overline \Omega}{}}

\newcommand{\bxi}{{\bar\xi}}

\newcommand{\bpsi}{{\bar\psi}{}}

\newcommand{\blambda}{{\bar\lambda}}

\newcommand{\bu}{{\bar u}}
\newcommand{\bq}{{\bar q}}

\newcommand{\bnabla}{{\overline \nabla}}

\newcommand{\cE}{{ {\cal E}   }}

\newcommand{\be}{\begin{equation}}
\newcommand{\ee}{\end{equation}}
\newcommand{\bea}{\begin{eqnarray}}
\newcommand{\eea}{\end{eqnarray}}

\newcommand{\ba}{\begin{array}} \newcommand{\ea}{\end{array}}

\def\im{{\rm i}}

\newcommand{\nn}{\nonumber}

\begin{document}

\begin{flushright}
ITP-UH-01/16
\end{flushright}\vspace{2cm}

\begin{center}
{\LARGE\bf Higher-derivative superparticle in AdS$_3$ space}
\end{center}
\vspace{1cm}

\begin{center}
{\Large\bf Nikolay Kozyrev${}^{a}$, Sergey Krivonos${}^{a}$ and Olaf Lechtenfeld${}^{b}$
}
\end{center}

\vspace{1cm}

\begin{center}
${}^a$ {\it
Bogoliubov  Laboratory of Theoretical Physics, JINR,
141980 Dubna, Russia}

\vspace{0.5cm}

${}^b$ {\it 
Institut f\"ur Theoretische Physik and Riemann Center for Geometry and Physics \\
Leibniz Universit\"at Hannover,
Appelstrasse 2, D-30167 Hannover, Germany}

\vspace{0.5cm}

{\tt  nkozyrev, krivonos@theor.jinr.ru, lechtenf@itp.uni-hannover.de}

\end{center}
\vspace{3cm}

\begin{abstract}\noindent
Employing the coset approach we construct component actions for a superparticle moving
in AdS$_3$ with $N{=}(2,0)$, $D{=}3$ supersymmetry partially broken to $N{=}2$, $d{=}1$.
These actions may contain higher time-derivative terms, which are chosen to possess
the same (super)symmetries as the free superparticle. In terms of the nonlinear-realization
superfields, the component actions always take a simpler form when written in terms of
covariant Cartan forms. We also consider in detail the reduction to the nonrelativistic case
and construct the corresponding action a Newton-Hooke superparticle and its higher-derivative
generalizations. The structure of these higher time-derivative generalizations is completely fixed
by invariance under the supersymmetric Newton-Hooke algebra extended by two central charges.
\end{abstract}

\newpage
\setcounter{page}{1}
\setcounter{equation}{0}
\section{Introduction}
The standard action of any particle moving in a flat spacetime is invariant under the target-space Poincar\'e group realized in a spontaneously broken manner. The spontaneously broken translations, orthogonal to the world-line of particle, and the Lorentz boosts rotating these translations into world-line translations,
give rise to Goldstone bosons, which appear in the particle actions. Usually, not all of these Goldstone bosons are independent of one another, and so there are
additional constraints reducing the number of independent fields to a set describing the physical degrees of freedom. In the supersymmetric case
the situation is more complicated, because in extended supersymmetry additional covariant constraints selecting irreducible supermultiplets
have to be found. These tasks can be algorithmically solved by using the nonlinear-realization (or coset) approach \cite{coset1}, suitably modified
for supersymmetric spacetime symmetries \cite{coset2}. In this approach, the corresponding constraints are conditions on the
Cartan forms (or on their $\theta$-components). Nevertheless, the coset approach fails to reproduce the superspace actions, because the superparticle Lagragian
is only {\it quasi-invariant\/} with respect to the super-Poincar\'e group and, therefore, it cannot be constructed in terms of Cartan forms. However, by passing to
component actions and focussing on the broken supersymmetry only, one may easily construct an ansatz for the invariant action in terms of $\theta=0$
projections of the Cartan forms. This program has been performed in our paper \cite{kkln} for a superparticle moving in flat $D=1{+}2$ spacetime with
$N{=}4$ supersymmetry partially broken to $N{=}2$. Moreover, the possible higher time-derivative terms, possessing the same symmetry as the free superparticle,
can also be constructed in terms of the Cartan forms. The role of unbroken supersymmetry is just to fix some free coefficients in the component action.

In the present paper, we investigate the application of the coset approach to a superparticle moving on AdS$_3$.
This leads to the following complications:
\begin{itemize}
\item One has to choose a suitable parametrization of the coset space (i.e.\ of the set of the physical bosonic fields), which may, even in the bosonic
case, drastically simplify the resulting actions;
\item The superspace constraints have to be properly covariantized, selecting irreducible supermultiplets which are rather distinct
from the flat-spacetime case;
\item The fermionic components should be defined such as to render the resulting Lagrangians readable.
All such choices are related by nonlinear invertible fields redefinitions \cite{SKuz}, but an improper choice may result
in highly complicated Lagrangians.
\end{itemize}
In the following we will solve all these tasks for a superparticle moving on AdS$_3$, with $N{=}(2,0)$, $D{=}3$ supersymmetry partially broken to $N{=}2$, $d{=}1$.
We will construct the corresponding component actions in terms of the $\theta=0$ projections of the Cartan forms and prove their invariance under
the $N{=}(2,0)$ AdS$_3$ algebra. The higher time-derivative terms share this symmetry, and the one with minimal higher derivatives
(the anyonic action) can also be written in terms of Cartan forms. The analysis will then be extended by the
nonrelativistic limit, which radically simplifies everything: The AdS$_3$ superparticle reduces to the
Newton-Hooke superparticle \cite{Anton1, BGKPRZ}, while the higher derivative terms acquire quite a compact form.

\setcounter{equation}{0}
\section{$N=(2,0)$ AdS$_3$ algebra and fixing the basis}
The action we are going to construct corresponds to the partial spontaneous breaking of $N=(2,0)$ AdS$_3$ supersymmetry. To start with, let us define
the $N=(2,0)$ AdS$_3$ super algebra in a standard way as (see e.g. \cite{Kuz1})
\bea\label{ads3}
&&\left[ M_{ab}, M_{cd}\right] = \epsilon_{ac} M_{bd}+\epsilon_{bd} M_{ac}+\epsilon_{ad}M_{bc}+\epsilon_{bc}M_{ad}\equiv \left(M\right)_{ab,cd}, \nn \\
&& \left[ M_{ab}, P_{cd}\right] = \left(P\right)_{ab,cd}, \quad \left[ P_{ab}, P_{cd}\right] = -\frac{m^2}{16} \left(M\right)_{ab,cd}, \nn \\
&& \left[ M_{ab}, Q_{c}\right] = \epsilon_{ac} Q_{b}+\epsilon_{bc} Q_{a} \equiv \left(Q\right)_{ab,c},\;
\left[ M_{ab}, \bQ_{c}\right] =  \left(\bQ\right)_{ab,c}, \nn \\
&&  \left[ P_{ab}, Q_{c}\right] =\im \frac{m}{4}  \left(Q\right)_{ab,c}, \;
\left[ P_{ab}, \bQ_{c}\right] =\im \frac{m}{4}  \left(\bQ\right)_{ab,c}, \quad
 \left[J,Q_a\right]=Q_a, \quad \left[J,\bQ_a\right]=-\bQ_a , \nn \\
&& \left\{ Q_a, \bQ_b \right\} = 2 P_{ab}+\im \frac{m}{2} M_{ab}+ \im m \epsilon_{ab} J.
\eea
Here, the generators $M_{ab}=M_{ba}, P_{ab}=P_{ba},\; a,b=1,2$ form the bosonic AdS$_3$ algebra while the fermionic generators $Q_a, \bQ_a$ together with the
$U(1)$ generator $J$ extend it to the $N=(2,0)$ AdS$_3$ one.

Note, that in our basis these generators obey the following conjugation rules
\be
\left(M_{ab}\right)^\dagger = - M_{ab},\;\left(P_{ab}\right)^\dagger = P_{ab},\;\left(J\right)^\dagger =J, \quad
\left( Q_a\right)^\dagger = \bQ_a.
\ee

To have close relations with the previously considered case of super particle moving  in  three-dimensional Poincar\'e space-time \cite{kkln}, one has to choose
the generators spanning $N=2, d=1$ super Poincar\'e algebra to which the AdS$_3$ supersymmetry will be broken to. One may easily check that if we  define
the generators $\left\{ P, Q, \bQ\right\}$ as
\be\label{newbasis1}
Q=Q_1+\im Q_2,\; \bQ=\bQ_1-\im \bQ_2,\quad P=P_{11}+P_{22}+\im \frac{m}{4}\left( M_{11}+M_{22} \right)+m J,
\ee
then they will form the $N=2, d=1$ super Poincar\'{e} algebra
\be\label{N2sP}
\left\{Q,\bQ\right\}= 2P,\qquad \left\{Q,Q\right\}=\left\{\bQ,\bQ\right\} =\left[ P,Q\right] =\left[P, \bQ\right]=0.
\ee
The remaining bosonic generators, having the proper form in the flat limit, may be defined as follows:
\bea\label{newgen}
&& P, \quad Z=P_{11}-P_{22}-2 \im P_{12}+\im \frac{m}{4} \left( M_{11}-M_{22}-2 \im M_{12} \right),\; \bZ= \left(Z\right)^\dagger, \nn \\
&& J_3 = \frac{\im}{4} \left( M_{11}+M_{22}\right), \quad T =\frac{\im}{4}\left( M_{11}-M_{22} -2 \im M_{12}\right), \;
\bT=\left( T\right)^\dagger.
\eea

Thus, the bosonic part of the algebra \p{ads3}, i.e. the algebra $so(2,2) \times u(1)$, acquires the form:
\bea\label{bosAdS}
&& \left[J_3, T\right]= T, \; \left[J_3, \bT\right]= -\bT, \; \left[T, \bT\right]= -2 J_3 , \nn \\
&& \left[ P, Z\right]= 2 m Z, \; \left[P, \bZ\right]= - 2 m \bZ, \; \left[Z, \bZ\right]= -4 m P +4 m^2 J , \nn \\
&& \left[J_3, Z\right] =Z, \; \left[ J_3, \bZ\right]= - \bZ , \nn \\
&& \left[T, P\right] =-Z, \; \left[\bT,P\right] =\bZ,\quad
\left[ T, \bZ\right]= - 2 P +2 m J,\;  \left[ \bT, Z\right]=  2 P - 2 m J .
\eea
Clearly, the relations \p{bosAdS} are maximally similar to the $D=3$ Poincar\'{e} ones we used in \cite{kkln} and go
to them in the limit $m=0$ (with decoupled generator $J$, of course).

As concerning the fermionic part of $N=(2,0)$ AdS$_3$ superalgebra \p{ads3},  it is natural to define the generators of broken supersymmetry as
\be\label{bsusygens}
S = \bQ_1 + \im \bQ_2, \quad \bS = Q_1 - \im Q_2.
\ee
Then commutation relations, which include spinor generators, read
\bea\label{AdS20}
&\left\{ Q, \bQ \right\} =2P, \;  \left\{ S, \bS \right\} =2P -4 m J, \; \left\{ Q, S \right\} =2\bZ, \; \left\{ \bQ, \bS\right\} =2Z,& \nn \\
&\left[ Z, Q\right] = -2m \bS, \; \left[ \bZ, \bQ\right] = 2m S, \; \left[ Z, S \right] = -2m \bQ, \; \left[ \bZ, \bS \right] = 2m Q,& \nn \\
&\left[ P, S\right] = -2m S, \; \left[ P, \bS\right] = 2m \bS, &\\
&\left[ T,Q  \right]=-\bS, \; \left[ \bT, \bQ\right] = S, \; \left[ T, S\right] = - \bQ, \; \left[ \bT, \bS \right] = Q,&\nn \\
&\left[ J_3, Q\right] = -\frac{1}{2}Q, \; \left[ J_3, \bQ\right] = \frac{1}{2}\bQ, \; \left[ J_3, S\right] = -\frac{1}{2}S, \; \left[ J_3, \bS\right] = \frac{1}{2}\bS,&\nn \\
&\left[ J, Q\right] = Q, \; \left[ J, \bQ\right] = -\bQ, \; \left[ J, S\right] = -S, \; \left[ J, \bS\right] = \bS.&\nn
\eea

\setcounter{equation}{0}
\section{Cartan forms and transformation properties}
In  the coset approach \cite{coset1,coset2}, the spontaneous breakdown of $S, \bS$ supersymmetry and $Z, \bZ$ translations is reflected in the structure of the coset element
\be\label{coset}
g = e^{\im t P} e^{\theta Q + \bar\theta \bQ} e^{\mpsi S + \mbpsi \bS} e^{\im \left( \mmU Z + \mmbU \bZ \right)} e^{\im \left( \mLambda T + \mbLambda \bT \right)}.
\ee
The $N=2$ superfields $\mmU(t,\theta,\bar\theta), \mpsi(t,\theta,\bar\theta)$ and $\mLambda(t,\theta,\bar\theta)$ are Goldstone superfields accompanying the
$N=(2,0)$ AdS$_3$ symmetry to $N=2, d=1$ super-Poincar\'e $\times U(1)^2$ breaking\footnote{These two additional $U(1)$ groups are formed by the generators $J$ and $J_3$.}. The transformation properties of the coordinates and the superfields are induced
by the left multiplication of the coset element \p{coset}
$$ g_0 \; g = g' \; h, \quad h \sim e^{\im \alpha J} e^{\im \beta J_3}.$$
The most important transformations read
\begin{itemize}
\item Unbroken SUSY $\left( g_0 = e^{\epsilon Q+ \bar\epsilon \bQ}\right)$
\be\label{Qsusytr}
\delta \theta = \epsilon, \; \delta \bar\theta =\bar\epsilon, \quad \delta t = \im \left( \epsilon \bar\theta+ \bar\epsilon \theta\right).
\ee
\item Broken SUSY $\left( g_0 = e^{\varepsilon S + \bar\varepsilon \bS}\right)$
\bea\label{Ssusytr}
&&\delta_S \theta = 4 m \tilde\varepsilon \mbpsi \theta ,\quad
\delta_S t = \im \left( \tilde\varepsilon \mbpsi + \bar{\tilde\varepsilon} \mpsi   \right)\big( 1- 6 m \theta\bar\theta  \big) -4 m \big( \tilde\varepsilon \theta \mmu - \bar{\tilde\varepsilon} \bar\theta \mmbu    \big)(1-2 m \mpsi\mbpsi),  \\
&& \delta_S \mpsi = \tilde{\varepsilon}\big( 1 -2m \theta\bar\theta \big)\big(1 +2 m \mpsi \mbpsi \big) - 8 \im m^2 \mpsi \big( \tilde{\varepsilon}\theta \mmu - \bar{\tilde\varepsilon}\bar\theta \mmbu  \big),\quad \delta_S \mmu = 2\im \bar{\tilde\varepsilon} \bar\theta \big( 1 -2 m \mpsi\mbpsi  \big)\big( 1  - 4 m^2 \mmu\mmbu   \big),\nn
\eea
where $\tilde\varepsilon = e^{2\im m t}\, \varepsilon $.
\item $Z, \bZ$-transformations $\left( g_0 = e^{\im (bZ + \bar b \bZ)}\right)$
\bea\label{Ztransf}
&& \delta_Z \theta =2 \im m \bar{\tilde b} \mpsi \left( 1 + 2 m \theta \bar\theta\right),\;
\delta_Z t = 4 m \left( {\tilde b} \theta \mpsi - \tilde{\bar b} \bar\theta \mbpsi\right)-
2\im m \left( {\tilde b} \mmbu -\bar{\tilde b} \mmu \right)\left(1+2 m \theta \bar\theta\right) \left( 1 -2 m \mpsi \mbpsi\right),
\nn \\
&&\delta_Z \mmu = {\tilde b} \left(1 - 4 m^2 \mmu\mmbu\right) \left(1+2 m \theta\bar\theta\right) \left( 1- 2m \mpsi\mbpsi\right),\nn\\
&&\delta_Z \mpsi =  2 \im m \bar{\tilde b} \bar\theta \left(1 +2 m \mpsi \mbpsi\right) +4 m^2 \mpsi \left( {\tilde b} \mmbu - \bar{\tilde b} \mmu\right) \left(1+2 m \theta \bar\theta\right).
\eea
where ${\tilde b} = e^{-2\im m t}\, b$.
\end{itemize}
Here, the coordinates of  stereographic projections were introduced
\be\label{stereo}
\mmu = \frac{ \tanh \left(2 m \sqrt{\mmU\; \mmbU}\right)}{2 m \sqrt{\mmU\; \mmbU}} \mmU, \qquad \mlambda = \frac{ \tanh \left(\sqrt{\mLambda \mbLambda}\right)}{\sqrt{\mLambda \mbLambda}}\mLambda.
\ee

The local geometric properties of the system are specified by the left-invariant Cartan forms
\be\label{susyCF}
g^{-1}dg = \im \Omega_P P + \im \Omega_Z Z + \im \bOmega_Z \bZ + \im \Omega_T T + \im \bOmega_T \bT +
 \im \Omega_3 J_3 + \im \Omega_J J  + \Omega_Q Q + \bOmega_Q \bQ + \Omega_S S + \bOmega_S \bS,
\ee
which look much more complicated than in the flat space-time $(m \rightarrow 0)$ \cite{kkln}
\bea\label{CF1}
&& \Omega_P = \frac{1}{1-\mlambda\mblambda}\left[ \left(1+\mlambda\mblambda\right) {\hat\Omega}_P - 2 \im \left( \mblambda {\hat\Omega}_Z-\mlambda  {\hat{ \bOmega}}_Z\right)\right], \quad \Omega_Q = \frac{\hat\Omega_Q - \im \mblambda \hat{\bOmega}_S}{\sqrt{1-\mlambda \mblambda}},
\nn \\[2mm]
&& \Omega_Z = \frac{1}{1-\mlambda\mblambda}\left[{\hat\Omega}_Z -\mlambda^2 {\hat{ \bOmega}}_Z +\im \mlambda {\hat\Omega}_P\right],\;\Omega_J = {\hat\Omega}_J -\frac{1}{1-\mlambda\mblambda}\left[ 2 m \mlambda\mblambda {\hat\Omega}_P - 2 \im m \left( \mblambda {\hat\Omega}_Z-\mlambda {\hat{\bOmega}}_Z\right)\right], \nn \\[2mm]
&& \Omega_T = \frac{d\mlambda}{1-\mlambda\mblambda}, \quad
\Omega_3 = \im\frac{\mlambda d\mblambda-d\mlambda \mblambda}{1-\mlambda\mblambda},  \quad \Omega_S = \frac{\hat\Omega_S - \im \mblambda \hat{\bOmega}_Q}{\sqrt{1-\mlambda \mblambda}},
\eea
where the hatted forms read
\bea\label{hatCF}
\hat{\Omega}_P &=& \frac{\left( 1+4m^2 \mmu\mmbu  \right)\hat\triangle t +2\im m \left( \mmu d\mmbu  - \mmbu d\mmu  \right) +8m \left(   \mmu\mpsi d\theta - \mmbu \mbpsi d\bar\theta \right)}{1-4m^2 \mmu\mmbu}, \nn \\
\hat{\Omega}_Z &=& \frac{d\mmu +2\im m \mmu \hat\triangle t - 2\im \left( \mbpsi d\bar\theta - 4m^2 \mmu^2 \mpsi d\theta  \right)}{1-4m^2 \mmu\mmbu},\nn \\
\hat\Omega_J &=& -\frac{8m^3 \mmu\mmbu }{1-4m^2 \mmu\mmbu} \hat\triangle t +2\im m^2 \frac{\mmbu d\mmu - d\mmbu \mmu}{1-4m^2 \mmu\mmbu} - 8 \frac{m^2 \left( \mmu\mpsi d\theta - \mmbu \mbpsi d\bar\theta  \right)}{1-4m^2 \mmu\mmbu} \nn \\
&& -8 m^2 \left[ dt -\im \left( \theta d\bar\theta + \bar\theta d\theta\right) \right] \mpsi \mbpsi +2 \im m \left(  \mpsi d\mbpsi  + \mbpsi d\mpsi  \right) \nn \\
&=& 2 m \left[ dt -\im \left( \theta d\bar\theta + \bar\theta d\theta\right) \right] - m \hat\triangle t - m \hat\Omega_P, \nn \\
\hat\Omega_Q & = & \frac{\triangle \theta - 2\im m \mmbu \triangle \mbpsi}{\sqrt{1-4 m^2 \mmu\mmbu }}, \quad
\hat\Omega_S  =  \frac{\triangle \mpsi - 2\im m \mmbu \triangle \bar\theta}{\sqrt{1-4 m^2 \mmu\mmbu }}.
\eea
Here,
\bea\label{diff}
\hat\triangle t &=& \left(1+4m \mpsi\mbpsi   \right) \left[ dt -\im \left( \theta d\bar\theta + \bar\theta d\theta+\mpsi d\mbpsi  + \mbpsi d\mpsi\right) \right] \equiv
\left(1+4m \mpsi\mbpsi   \right) \triangle t,\nn \\
\triangle\theta & = & \left(  1 -2m \mpsi\mbpsi    \right)d\theta, \quad \triangle \mpsi = d\mpsi - 2\im m \mpsi \left[ dt -\im \left( \theta d\bar\theta + \bar\theta d\theta\right) \right] .
\eea

In what follows, we find it convenient to define the covariant derivatives similarly to the flat case, i.e.   with respect to differentials $\triangle t, d\theta, d\bar\theta$:
\be\label{cD0}
dt \frac{\partial}{\partial t}+
d\theta \frac{\partial}{\partial\theta} + d\bar\theta \frac{\partial}{\partial\bar\theta}=
\triangle t \nabla_t + d\theta \nabla +d\bar\theta \bnabla.
\ee
Explicitly, they read
\be\label{cD}
\nabla_t =E^{-1} \partial_t, \quad \nabla = D -\im \left( \mbpsi \nabla \mpsi + \mpsi \nabla\mbpsi\right) \partial_t,\quad
\bnabla=\bD -\im\left( \mbpsi \bnabla \mpsi + \mpsi \bnabla\mbpsi\right) \partial_t,
\ee
where
\bea\label{cD1}
 E= 1 + \im\left( \dot\mpsi \mbpsi +  \dot\mbpsi \mpsi\right), \qquad
 D= \frac{\partial}{\partial \theta } -\im \bar\theta \partial_t,\quad\bD= \frac{\partial}{\partial \bar\theta } -\im \theta \partial_t\; :  \quad  \left\{D, \bD \right\}=-2\im \partial_t.
\eea
These derivatives obey the following algebra,
\bea\label{CDcom}
&& \left\{\nabla, \bnabla\right\} =-2 \im\left( 1+  \nabla\mpsi\bnabla\mbpsi +
\bnabla\mpsi\nabla \mbpsi\right)  \nabla_t , \quad \left\{ \nabla, \nabla\right\} = -4 \im \nabla\mbpsi \nabla\mpsi\nabla_t, \quad
\left\{ \bnabla, \bnabla\right\} = -4 \im \bnabla\mbpsi \bnabla\mpsi\nabla_t, \nn \\
&& \left[ \nabla_t, \nabla\right] = - 2\im \left( \nabla \mbpsi \nabla_t \mpsi +\nabla \mpsi \nabla_t\mbpsi \right) \nabla_t, \quad
\left[ \nabla_t, \bnabla\right] = - 2\im \left( \bnabla \mbpsi \nabla_t \mpsi +\bnabla\mpsi \nabla_t\mbpsi \right) \nabla_t .
\eea

\setcounter{equation}{0}
\section{Preliminary consideration: the bosonic action}
Before considering the full supersymmetric AdS$_3$ system it makes sense to analyze its bosonic sector.

The bosonic sector of our $N=(2,0)$ supersymmetric AdS$_3$ superalgebra \p{ads3} contains the bosonic AdS$_3$ algebra (i.e. $so(2,2)$ algebra) commuting with the $U(1)$ algebra spanned by the generator $J$. In this subsection we are going to consider the spontaneous breakdown of this AdS$_3 \times S^1 $  symmetry down to
$d=1 \mbox{ Poincar\'{e} } \times U(1)^2$ algebra, generated by $P, J$ and $J_3$ generators. Therefore,  our coset element is just the $(\theta,\psi \rightarrow 0)$ limit of the full coset element \p{coset}
\be\label{boscoset}
g= e^{\im t P}\;  e^{\im\left( U Z + {\overline  U} \bZ\right)} \; e^{\im\left( \Lambda T + \bLambda \bT\right)}.
\ee
The corresponding $(\theta,\psi \rightarrow 0$) limit of the Cartan forms read
\bea\label{bosCF1}
&& \omega_P = \frac{1}{1-\lambda\blambda}\left[ \left(1+\lambda\blambda\right) {\hat\omega}_P - 2 \im \left( \blambda {\hat\omega}_Z-\lambda  {\hat{ \overline\omega}}_Z\right)\right],\quad  \omega_Z = \frac{1}{1-\lambda\blambda}\left[{\hat\omega}_Z -\lambda^2 {\hat{ \overline\omega}}_Z +\im \lambda {\hat\omega}_P\right],\nn  \\[2mm]
&& \omega_J = {\hat\omega}_J -\frac{1}{1-\lambda\blambda}\left[ 2 m \lambda\blambda \omega_P - 2 \im m \left( \blambda {\hat\omega}_Z-\lambda {\hat{\overline \omega}}_Z\right)\right], \quad \omega_T = \frac{d\lambda}{1-\lambda\blambda},  \quad
\omega_3 = \im\frac{\lambda d\blambda-d\lambda \blambda}{1-\lambda\blambda},
\eea
where
\bea\label{bosCF2}
&& {\hat\omega}_P = \frac{1}{1-4 m^2 u \bu}\left[ \left(1+4 m^2 u \bu\right) dt +2 \im m \left( u d \bu -\bu du\right) \right],\nn \\[2mm]
&& {\hat\omega}_Z = \frac{1}{1-4 m^2 u \bu}\left[ d u + 2 \im m u dt\right],\quad
{\hat{\overline\omega}}_Z = \frac{1}{1-4 m^2 u \bu}\left[ d \bu - 2 \im m \bu dt\right],\nn \\[2mm]
&& {\hat\omega}_J =-\frac{2 m^2}{1-4 m^2 u \bu}\left[ 4 m u \bu dt +\im \left( u d\bu -\bu d u\right)\right] = m \left( dt - {\hat\omega}_P\right).
\eea

To reduce the number of independent Goldstone fields, similarly to the flat space case \cite{kkln},  one may impose  the following conditions on the Cartan forms ${\omega}_Z$ and ${\bar\omega}_Z$ (inverse Higgs phenomenon \cite{ih}),
\bea\label{ih}
&&\omega_Z = 0 \; \Rightarrow \; \hat{\omega}_Z = -\im \frac{\lambda}{1+\lambda\blambda} \hat{\omega}_P\; \Rightarrow \; \frac{\lambda}{1+\lambda\blambda}=\im \frac{\dot{u}+2 \im m u}{1+4 m^2 u \bu +2 \im m( u \dot{\bu} -\bu \dot{u})},\nn \\
&&{\overline\omega}_Z = 0 \;\Rightarrow \;\hat{\overline\omega}_Z = \im \frac{\blambda}{1+\lambda\blambda} \hat{\omega}_P \; \Rightarrow\; \frac{\blambda}{1+\lambda\blambda}=-\im \frac{\dot{\bu}-2 \im m \bu}{1+4 m^2 u \bu +2 \im m( u \dot{\bu} -\bu \dot{u})},
\eea
and, therefore,
\be\label{lambdau}
\dot{u} = -\im \displaystyle\frac{(2 m u+\lambda)(1+2 m u \blambda)}{1+\lambda\blambda+ 2m (\bu \lambda+u \blambda)} .
\ee
These constraints are purely kinematic ones. Thus, to realize this spontaneous breaking of AdS$_3 \times U(1)$
symmetry we need one complex  scalar field, $u(t)$ and ${\bu}(t)$.

Using the constraints \p{ih}, one may further simplify the Cartan forms ${\omega}_P, \; {\omega}_J$ \p{bosCF1} to be
\be\label{omegaP}
{\omega}_P=\frac{1-\lambda\blambda}{1+\lambda\blambda}{\hat\omega}_P, \quad \omega_J = m dt -m \frac{1-\lambda\blambda}{1+\lambda\blambda}{\hat\omega}_P.
\ee

Clearly, the simplest action, invariant under full AdS$_3 \times U(1)$  symmetry, is
\bea\label{action1}
S_0=- m_0\int {\omega}_P &=& -m_0\int dt \sqrt{\left(1+ \frac{ 2\im m \left( \dot{u} \bu-u \dot{\bu}\right)}{1-4 m^2 u \bu}\right)^2 - 4  \frac{{\dot u}\dot{\bu}}{\left(1-4 m^2 u \bu\right)^2}} \nn \\
&=& -m_0 \int dt \frac{1-\lambda \blambda}{ 1+ \lambda \blambda +2m (\lambda \bu +\blambda u)}.
\eea
One may check, that the curvature of the space with the metric
\be
ds^2 =- \left(dt+ \frac{ 2\im m \left( d{u} \bu-u d{\bu}\right)}{1-4 m^2 u \bu}\right)^2 + 4
 \frac{{d u} d{\bu}}{\left(1-4 m^2 u \bu\right)^2}
\ee
is equal to ${\cal R}=-6 m^2$.

Keeping in  mind that the Cartan form $\omega_3$ is shifted by the full time derivative under all transformations of the AdS$_3 \times U(1)$ group, the invariant anyonic term, i.e. the action which results in the at most the third order time derivatives equations of motion of the fields and which possesses the invariance under full AdS$_3 \times U(1)$  symmetry,  acquires the form
\bea\label{anyonbosaction}
S_{anyon} = -\int \omega_3 = \im \int dt \frac{\dot \lambda\blambda - \dot\blambda \lambda}{1-\lambda\blambda}.
\eea
In terms of $u,\, \bu$ and their derivatives it reads
\bea\label{anyonu}
S_{anyon} &=& \int dt \left\{  \frac{2\im\left(\ddot u \dot \bu - \ddot \bu \dot u  \right) -  8m \dot u \dot \bu +8\im m^2 \left( \dot u \bu - \dot\bu u   \right)+ 4m\left( \ddot u \bu + \ddot\bu u  \right)}{\sqrt{\left( 1 - 4m^2 u\bu -2\im m (\dot \bu u - \dot u \bu)   \right)^2 - 4\dot u \dot\bu}} \right.  \nn \\
&& \left. \times \left[1 + 4m^2 u\bu +2\im m \left( u\dot\bu - \bu\dot u   \right) + \sqrt{\left( 1 - 4m^2 u\bu -2\im m (\dot \bu u - \dot u \bu)   \right)^2 - 4\dot u \dot\bu}    \right]^{-1}\right\}.
\eea

The  actions $S_0$ \p{action1} and $S_{anyon}$ \p{anyonu} may be slightly simplified by passing to new variables $q, {\bar q}$ defined as
\be\label{q}
q = e^{2\im m t} u,\; {\bar q} =e^{-2\im m t} \bu .
\ee
In terms of these variables the action of the AdS$_3$ particle reads
\be\label{actionq}
S_0 = -m_0 \int dt \sqrt{\left( 1 - 2\im m \frac{\dot q \bq - \dot\bq q}{1-4m^2 q \bq}      \right)^2 -  \frac{4 \dot q \dot \bq}{\left( 1-4m^2 q \bq  \right)^2}},
\ee
while the anyonic action \p{anyonu} is simplified to be
\bea\label{anyonq}
S_{anyon} &=& \int dt \left\{ \frac{2\im \left( \ddot q \dot\bq - \ddot \bq \dot q   \right) + 8 m \dot q \dot\bq}{\sqrt{\left(1-4m^2 q \bq + 2\im m(\dot \bq q - \dot q \bq) \right)^2 - 4 \dot q \dot \bq}}  \right. \nn \\
&&\left. \times \left[ 1 -4m^2 q \bq +2\im m(\dot \bq q - \dot q \bq) + \sqrt{\left(1-4m^2 q \bq + 2\im m(\dot \bq q - \dot q\bq) \right)^2 - 4 \dot q \dot \bq}       \right]^{-1} \right\}.
\eea
This action can be rewritten through the Lagrangian ${\cal L} _{PS2}$ of a particle on the pseudosphere  and  connection ${\cal A}_{\mu}$ as
\be\label{anyonqAL}
S_{anyon} = \int dt \frac{2\im \displaystyle\frac{\ddot q \dot \bq - \ddot \bq \dot q}{\left( 1- 4 m^2 q\bq \right)^2} + 8 m {\cal L}_{PS2} }{\sqrt{\left( 1 - {\cal A}_{\mu} \dot q^{\mu} \right)^2 - 4 {\cal L}_{PS2}} \left( 1 - {\cal A}_{\mu} \dot q^{\mu}+\sqrt{\left( 1 - {\cal A}_{\mu} \dot q^{\mu} \right)^2 - 4 {\cal L}_{PS2}} \right)}.
\ee
Here
\be\label{newdefs}
{\cal L} _{PS2} = \frac{\dot q \dot \bq}{\left( 1 - 4m^2 q \bq  \right)^2}, \qquad
 {\cal A}_{\mu} \dot q^{\mu}= 2\im m \frac{\dot q \bq - \dot\bq q}{1-4m^2 q \bq}.
\ee

The AdS$_3$ anyon action, written in terms of the $\lambda, \blambda$ variables  \p{anyonbosaction}, has the same form as
in the flat space-time case \cite{kkln}. This analogy is slightly broken  for the AdS$_3$ rigid particle action, which is invariant under the full AdS$_3 \times U(1)$  symmetry and leads to equations of motion of at most fourth order in time derivatives,
\be\label{rigid}
S_{rigid} = \beta \int \frac{\omega_T {\bar \omega}_T}{\omega_P} =\int dt \left( 1 +2 m \frac{ \lambda \bu + \blambda u}{1+\lambda \blambda}\right)
\left( \frac{1+\lambda \blambda}{\left(1 -\lambda \blambda\right)^3}\right) {\dot\lambda}{\dot\blambda}.
\ee
The  term proportional to $m$ is  needed to provide  invariance with respect to the AdS$_3$ symmetry, realized by  left
multiplications of the coset element \p{boscoset} as follows,
\bea\label{bostrlaws}
g_0 = e^{\im (aT + \bar a \bT)} &\Rightarrow& \delta_T t = -\im a \bar u\big( e^{2\im m t} +1  \big) + \im \bar a u \big( e^{-2\im m t} +1  \big), \quad \delta_T \lambda = a - \bar a \lambda^2, \nn \\
&& \delta_T u = \frac{a}{2m} e^{-2\im m t} \big( 1 - 4 m^2 u \bar u  \big) - \frac{1}{2m} \big( a -4 m^2 \bar a u^2   \big), \\
g_0 = e^{\im (bZ + \bar b \bZ)} & \Rightarrow & \delta_Z t = -2\im m \big( b\, e^{-2\im m t} \bar u - \bar b\, e^{2\im m t}u    \big), \quad \delta_Z u = b \, \big( 1 - 4 m^2 u\bar u   \big)e^{-2\im m t}, \quad \delta_Z \lambda =0.\nn
\eea

Let us finally stress that the  actions \p{action1}, \p{anyonu} \p{rigid} we constructed  are precisely the flat space-time expressions when expressed in the terms of Cartan forms \cite{kkln}.

\setcounter{equation}{0}
\section{Fully supersymmetric case}
To construct supersymmetric component actions, invariant under both unbroken $Q$ and broken $S$ supersymmetries, in full analogy with the flat case \cite{kkln}, one has to perform four steps:
\begin{itemize}
\item Impose some additional constraints to reduce the number of independent superfields and  impose
irreducibility constraints on the essential superfields;
\item Find the transformation properties of the physical components under both supersymmetries;
\item Write an ansatz for the component actions invariant under broken supersymmetries. The corresponding invariants are provided by the Cartan forms evaluated at $\theta=\bar\theta=0$ condition;
\item Fix the arbitrary parameters in the ansatz by demanding invariance under the
unbroken supersymmetry.
\end{itemize}
Let us go through these steps.

\subsection{Irreducibility conditions}
{}From the beginning, in our coset \p{coset} there are three independent complex superfields $\mmu, \mpsi,\mlambda$ (considering the redefinitions \p{stereo}). To reduce the number of independent superfields we impose the same conditions \p{ih} as in the bosonic sector,
\be\label{SUSYih}
\left\{
\begin{array}{l}
\Omega_Z =0 \\
\bOmega_Z =0
\end{array} \right. \quad \Rightarrow \quad
\hat\Omega_Z = - \im \frac{\mlambda}{1+\mlambda\mblambda}\hat\Omega_P, \; \hat\bOmega_Z = \im  \frac{\mblambda}{1+\mlambda\mblambda}\hat\Omega_P.
\ee
Equating the coefficients of the differentials $\triangle t, d\theta $ and $d \bar\theta$ we get
\bea\label{IC1}
&&\left( 1 - 4 m \mpsi \mbpsi\right) \nabla_t\mmu = -2\im m \mmu  - \im  \frac{\mlambda\big( 1-4 m^2 \mmu \mmbu  \big) }{1 +\mlambda\mblambda +2m \big( \mlambda\mmbu + \mblambda\mmu   \big)},\nn \\
&&\left( 1 - 4 m \mpsi \mbpsi\right)\nabla_t\mmbu = 2\im m \mmbu  + \im  \frac{\mblambda\big( 1-4 m^2 \mmu \mmbu  \big) }{1 +\mlambda\mblambda +2m \big( \mlambda\mmbu + \mblambda\mmu   \big)}
\eea
and
\bea
&& \nabla \mmu +4 m \mmu \mpsi \nabla_t \mmu = 0,\quad \bnabla \mmbu -4 m \mmbu \mbpsi \nabla_t \mmbu = 0, \label{IC21} \\
&& \bnabla \mmu = - 2\im \mbpsi \left(  1-4 m^2 \mmu \mmbu  +2\im m  \mmbu\nabla_t \mmu\right), \;
\nabla \mmbu = - 2\im \mpsi \left(  1-4 m^2 \mmu \mmbu  +2\im m  \mmu\nabla_t \mmbu\right). \label{IC22}
\eea
These relations simplify the form  $\Omega_P$ to
\be\label{OmegaP}
\Omega_P =\frac{1-\mlambda \mblambda}{1+\mlambda \mblambda}\hat\Omega_P,\quad
\hat\Omega_P = \left( 1 + 4 m \mpsi \mbpsi\right) \left[ \frac{ \triangle t +4 m \left( \mmu \mpsi d\theta -\mmbu \mbpsi d\bar\theta\right)}{1 +2 m \frac{\mmbu \mlambda +\mmu \mblambda}{1+\mlambda \mblambda}} \right].
\ee

In principle,  \p{IC1}, \p{IC21}, \p{IC22} solve all tasks. Indeed, using \p{IC1} one may
express the superfields $\mlambda, \mblambda$ in terms of  time derivatives of  $\mmu,\mmbu$, while \p{IC21} can be solved to express the fermionic superfields  $\mpsi, \mbpsi$ in terms of spinor covariant
derivatives of the same $\mmu,\mmbu$. Thus, like in the flat case \cite{kkln}, we remain with only one $N=2$ complex bosonic superfield $\mmu(t, \theta, \bar\theta)$, restricted by  \p{IC21} to be
covariantly chiral, with slightly modified chirality conditions. However, in what follows we are going to use as
independent components the $\theta=\bar\theta=0$ projections of the superfields $\mpsi, \mbpsi$ instead of the
projections of $ \bnabla \mmu$ and $\nabla \mmbu$. Therefore, it would be
useful to find the consequences of the constraints \p{IC1}, \p{IC21}, \p{IC22}.

First of all, acting by $\nabla$ on the first equation in \p{IC21} and by $\bnabla$ on the second one and using the algebra \p{CDcom} of the covariant derivatives, we get the conditions
\be\label{IC31}
\nabla \mpsi +4 m \mmu \mpsi \nabla_t \mpsi = 0,\quad \bnabla \mbpsi -4 m \mmbu \mbpsi \nabla_t \mbpsi = 0.
\ee
Note, that this asserts the self-consistency of the modified chirality constraints \p{IC21} because
\be\label{prove1}
\left\{ \nabla +4 m \mmu \mpsi \nabla_t, \nabla +4 m \mmu \mpsi \nabla_t\right\}=0, \quad
\left\{ \bnabla  -4 m \mmbu \mbpsi \nabla_t, \bnabla  -4 m \mmbu \mbpsi \nabla_t\right\} =0.
\ee
Secondly, acting by $\bnabla$ on the first equation in \p{IC21} and by $\nabla$ on the first equation in \p{IC22} and adding the
results, after quite lengthly calculations with heavy use of \p{CDcom}, we obtain
\be\label{IC32}
\nabla\mbpsi
= - \im \frac{\mlambda +2m \mmu}{1 + 2m \mmbu \mlambda}\big( 1 - 4m\mpsi\mbpsi   \big) - 8\im m^2 \mmu \mpsi\mbpsi - 4m\mmu \mpsi\nabla_t \mbpsi.
\ee
Repeating similar calculations with the second equations in \p{IC21}, \p{IC22} yields the conjugated expression
\be\label{IC33}
\bnabla\mpsi
=  \im \frac{\mblambda +2m \mmbu}{1 + 2m \mmu \mblambda}\big( 1 - 4m\mpsi\mbpsi   \big) + 8\im m^2 \mmbu \mpsi\mbpsi + 4m\mmbu \mbpsi\nabla_t \mpsi.
\ee
Now we have all  ingredients needed for constructing the component action.

Before closing this subsection let us visualize a more simple way to obtain  \p{IC32} and \p{IC33}. The idea consists
in the using the constraints\footnote{These conditions, being  some variant of the superembedding conditions \cite{Dima},  were trivial in the flat case \cite{kkln}.}
\be\label{Alt1}
\Omega_S |_{\Omega_Q, \bOmega_Q} = 0,\quad \bOmega_S |_{\Omega_Q, \bOmega_Q} = 0.
\ee
Here, the notation ${|_{\Omega_Q, \bOmega_Q}}$ means that the Cartan forms $\Omega_S$ and $\bOmega_S$ must be expanded in
the forms $\Omega_P, \Omega_Q, \bOmega_Q$ before nullifying their ${|_{\Omega_Q, \bOmega_Q}}$-projections.
At first sight these conditions seem to be more complicated due to the highly nontrivial structure of the Cartan forms involved.
However,  this is not the case and the calculations can be simplified using the following procedure.

With the help of our definitions \p{CF1}, the first constraint in \p{Alt1} can be formally represented as
\be\label{Alt2}
\Omega_S = \frac{\hat\Omega_S - \im \mblambda \hat{\bOmega}_Q}{\sqrt{1-\mlambda \mblambda}} = \Omega_P \; X
\ee
where $X$ is  some expression which is defined by this equation. The main difference between \p{Alt1} and
\p{Alt2} is that the latter one is written as the equation on  forms. Therefore, one may just substitute in \p{Alt2}
the exact expressions from \p{hatCF} and equate on both sides the coefficients of the differentials $\triangle t, d \theta$
and $d\bar\theta$. The $\triangle t$ coefficient relation yields $X$ in terms of $\nabla_t \mpsi$.
Substituting this expression for $X$ in the $d\theta, d\bar\theta$ -projections of \p{Alt2} we immediately
obtain the first equation in \p{IC31} and also \p{IC33}. The same procedure applied to the second constraint in \p{Alt1} produces the second equation in \p{IC31} as well as \p{IC32}.

These considerations demonstrate that the constraints \p{Alt1} are a consequence of our basic constraints \p{SUSYih}.

\subsection{Transformation properties of the components}
As we are going to construct the component actions, we need to know the transformation laws for the components. We denote the components of the superfields in the following way,
\be\label{def2}
\mmu|_{\theta=0} =u,\; \mmbu|_{\theta=0} = \bu,\quad
\mpsi|_{\theta=0} =\psi,\; \mbpsi|_{\theta=0} = \bpsi,\quad
\mlambda|_{\theta=0} =\lambda,\; \mblambda|_{\theta=0} = \blambda.
\ee
The equations \p{IC1} evaluated at $\theta=\bar\theta=0$ provide  relations between $\lambda, \blambda$
and the time derivatives of $u, \bu$. Thus, $\lambda, \blambda$ are not independent components. We introduce these variables just
to simplify many expressions in what follows.

\noindent{\it Broken $S$ supersymmetry}\\
The transformation properties of our components \p{def2} under broken supersymmetry can be easily learned from \p{Ssusytr}. Before listing these transformations, we point out that, in contrast to the flat case \cite{kkln}, the superspace
coordinates $\theta$ and $\bar\theta$ are not invariant under the broken supersymmetry \p{Ssusytr},
$$
\delta_S \theta = 4 m \varepsilon\, e^{2\im m t } \mbpsi \theta,\quad  \delta_S \bar\theta = -4 m \bar\varepsilon\, e^{-2\im m t} \mpsi \bar\theta.
$$
However, the right hand sides of these variations disappear in the limit $\theta=\bar\theta=0$ and, thus, the set of the components $\left\{ u, \bu, \psi, \bpsi\right\}$ is closed under the broken supersymmetry. The corresponding transformations read
\bea\label{Scomptr}
&&\delta_S t =\im \left( \varepsilon\, e^{2\im m t } \bpsi + \bar\varepsilon\, e^{-2\im m t} \psi \right), \nn\\
&&\delta_S \psi = \varepsilon\, e^{2\im m t } \left( 1 + 2m \psi\bpsi   \right),\;
\delta_S \bpsi = \bar\varepsilon\, e^{-2\im m t } \left( 1 + 2m \psi\bpsi   \right),
  \quad \delta_S u =0,\;  \delta_S \bu =0.
\eea
It is rather easy to check that the expression
\be\label{newdt}
\left( 1 + 4 m \psi \bpsi\right) \triangle t |_{\theta=0} =\left( 1 + 4 m \psi \bpsi\right)\left[ dt - \im \left( \psi d\bpsi +
\bpsi d \psi\right)\right]\equiv \cE dt
\ee
is invariant with respect to \p{Scomptr}. Therefore, it is natural to define a new covariant derivative as
\be\label{ccD}
\cD_t = \cE^{-1} \partial_t, \quad \cE^{-1} = \left( 1-4m \psi\bpsi\right)\left[1 -\im \left( \cD_t \psi \bpsi + \cD_t \bpsi \psi    \right)\right].
\ee
It then immediately follows from \p{newdt} and \p{ccD} that
\be\label{trdtu}
\delta_S \cD_t u=0, \qquad \delta_S \cD_t \bu =0 \; \Rightarrow \; \delta_S \lambda = \delta_S \blambda =0.
\ee

\noindent{\it Unbroken $Q$ supersymmetry}\\
The  transformations under unbroken $Q$-supersymmetry can be defined in a usual way as
\bea\label{Qtrgen}
\delta_Q f &=& - \big( \epsilon D + \bar\epsilon\bD   \big)\mf |_{\theta\rightarrow 0} =  - \big( \epsilon \nabla + \bar\epsilon\bnabla  \big)\mf |_{\theta\rightarrow 0} - H \partial_t, \nn \\
H &=& \im \epsilon \big( \mpsi \nabla\mbpsi  + \mbpsi \nabla\mpsi \big) |_{\theta\rightarrow 0} + \im \bar\epsilon \big( \mpsi \bnabla\mbpsi  + \mbpsi \bnabla\mpsi \big) |_{\theta\rightarrow 0}.
\eea
For example,
\be\label{Qtr1}
\delta_Q u = 2 \im \bar\epsilon \bpsi \big( 1 -4 m^2 u \bar u   \big) +4 m \big( \epsilon u  \psi -  \bar\epsilon \bar u  \bar \psi   \big)\cD_t u - H \dot u, \quad \delta_Q \psi = -  \big( \epsilon \nabla\mpsi + \bar\epsilon\bnabla\mpsi  \big) |_{\theta\rightarrow 0} - H \dot\psi.
\ee
Another important object is the vielbein  $\cE$ \p{ccD} which transforms as follows
\bea
\delta_Q \cE & =&  2\im \big( 1 + 4 m \psi\bpsi   \big) \cE \left[ \big( \epsilon \nabla\mpsi + \bar\epsilon\bnabla\mpsi     \big)|_{\theta\rightarrow 0} \dot\bpsi + \big( \epsilon \nabla\mbpsi + \bar\epsilon\bnabla\mbpsi     \big)|_{\theta\rightarrow 0} \dot\psi   \right]\nn \\
&& -4 m \cE \big( \epsilon \nabla\mpsi + \bar\epsilon\bnabla\mpsi     \big)|_{\theta\rightarrow 0}\bpsi + 4m \cE \big( \epsilon \nabla\mbpsi + \bar\epsilon\bnabla\mbpsi     \big)|_{\theta\rightarrow 0}\psi - \partial_t \big( H \cE   \big).\label{QtrE0}
\eea
Of course, to find the explicit form of the transformations \p{QtrE0} one has to use the relations \p{IC31}, \p{IC32}, \p{IC33} evaluated at $\theta=\bar\theta=0$:
\bea
&& \left(\nabla \mpsi\right)|_{\theta=0}  +4 m u \psi \cD_t \psi = 0,\quad \left(\bnabla \mbpsi\right)|_{\theta=0} -4 m \bu \bpsi \cD_t \bpsi = 0, \nn \\
&& \left(\nabla\mbpsi\right)|_{\theta=0}
= - \im \frac{\lambda +2m u}{1 + 2m \bu \lambda}\big( 1 - 4m\psi\bpsi   \big) - 8\im m^2 u \psi\bpsi - 4m u \psi\cD_t \bpsi, \nn\\
&& \left( \bnabla\mpsi\right)|_{\theta=0}
=  \im \frac{\blambda +2m \bu}{1 + 2m u \blambda}\big( 1 - 4m\psi\bpsi   \big) + 8\im m^2 \bu \psi\bpsi + 4m\bu \bpsi\cD_t \psi.
\label{Qtr2}\eea
In particular, the transformation \p{QtrE0} acquires the form
\be\label{QtrE1}
\delta_Q  \cE = - \partial_t \big( H \cE \big) + 2 \cE \frac{\lambda+2 m u}{1+2m \bar u \lambda}\epsilon\big(\cD_t \psi - 2\im m \psi \big) - 2 \cE \frac{\bar\lambda+2 m \bar u}{1+2m u \bar \lambda}\bar\epsilon\big(\cD_t \bpsi +2\im m \bpsi \big).
\ee
Finally, we stress that the relations between the components $u, \bu$ and $\lambda, \blambda$ are given by the following
expressions,
\be\label{Qtrdu}
 \cD_t u = -2\im m u  - \im  \frac{\lambda\big( 1-4 m^2 u \bu  \big) }{1 +\lambda\blambda +2m \big( \lambda\bu + \blambda u   \big)},\quad
\cD_t\bu = 2\im m \bu  + \im  \frac{\blambda\big( 1-4 m^2 u \bu  \big) }{1 +\lambda\blambda +2m \big( \lambda\bu + \blambda u   \big)}
\ee
\subsection{Actions}
We are ready to construct the supersymmetric generalization of the actions \p{action1} and \p{anyonbosaction}. As they have
different dimension, these actions must be invariant individually.\vspace{0.25cm}

\noindent{\bf Superparticle}\vspace{0.25cm}

It is easy to check that the evident ansatz
$$
\int dt\; \cE F(u, \bu, \lambda, \blambda)
$$
is  invariant under the broken supersymmetry for any function $F$ because, in virtue of \p{Scomptr}, \p{newdt}, \p{trdtu},
\be\label{trAct0}
\delta_S F =0, \quad \delta_S \left( dt \cE\right) =0.
\ee
The desired bosonic limit \p{action1} immediately fixes the function $F$ up to constant $\alpha$
\be\label{SusyAct0}
S_0 =-m_0 \int dt\; \cE \left[ \alpha + \frac{1-\lambda\blambda}{1 +\lambda\blambda+ 2m \big(u\blambda + \bu \lambda \big)}   \right].
\ee
This constant $\alpha$ can be determined as unity  either from linearized $Q$ supersymmetry invariance or from the flat space-time action of \cite{kkln}. Let us explicitly demonstrate that the action
\be\label{SusyAct1}
S_0 =-m_0 \int dt\; \cE \left[ 1 + \frac{1-\lambda\blambda}{1 +\lambda\blambda+ 2m \big(u\blambda + \bu \lambda \big)}   \right]
\equiv -m_0 \int dt {\cal L}
\ee
is invariant under the unbroken $Q$ supersymmetry.

Using \p{Qtr1} and \p{Qtrdu}, one  finds that
\be\label{Qtrlambda}
\delta_Q \lambda = -2\bar\epsilon \big( \cD_t \bpsi + 2\im m \bpsi  \big)\frac{1+ 2m \bar u \lambda}{1+2 m u \bar\lambda} \big( 1 +\lambda\bar\lambda + 2m (u\bar\lambda - \bar u \lambda)   \big) + 4m  \big( \epsilon u \psi - \bar\epsilon \bar u \bpsi  \big)\cD_t \lambda - H \partial_t \lambda, \; \delta_Q \blambda = \left( \delta_Q \lambda\right)^\dagger.
\ee
Now, the variation of integrand in \p{SusyAct1} reads
\bea\label{Qlambdalagrtr1}
\delta_Q {\cal L}  = - \partial_t \big( H {\cal L}  \big) + 4m \cE \big( \epsilon u \psi - \bar\epsilon \bar u \bpsi   \big)\cD_t \left[  \frac{1-\lambda\bar\lambda}{1 +\lambda\bar\lambda+ 2m \big(u\bar\lambda + \bar u \lambda \big)}  \right]\nn\\
 +4 m \cE \big( \epsilon \cD_t \psi u - \bar\epsilon \cD_t \bpsi \bar u   \big)\frac{1 - \lambda \bar\lambda}{1+\lambda \bar\lambda +2m \big( u\bar\lambda + \bar u\lambda   \big) }  \nn \\
-4 \im m \cE \epsilon\psi \frac{\big(1 -\lambda\bar\lambda \big)\big( 1+ 2m  u\bar\lambda  \big)\big(\lambda + 2 m u)}{\big[1 +\lambda\bar\lambda+ 2m \big(u\bar\lambda + \bar u \lambda \big) \big]^2} - 4 \im m \cE \bar \epsilon\bpsi \frac{\big(1 -\lambda\bar\lambda \big)\big( 1+ 2m  \bar u\lambda  \big)\big(\bar\lambda + 2 m \bar u)}{\big[1 +\lambda\bar\lambda+ 2m \big(u\bar\lambda + \bar u \lambda \big) \big]^2}.
\eea
Using the relations \p{Qtrdu}, the last line in \p{Qlambdalagrtr1} may be represented as
$$
4 m \cE \big( \epsilon \psi \cD_t u - \bar\epsilon \cD_t \bpsi \cD_t \bu   \big)\frac{1 - \lambda \blambda}{1+\lambda \blambda +2m \big( u\blambda + \bu\lambda   \big) },
$$
and, therefore,
\be\label{QED1}
\delta_Q {\cal L} = - \partial_t \big( H {\cal L}  \big) + 4m \cE \cD_t \left[    \frac{\big( \epsilon \psi u - \bar\epsilon\bar\psi \bar u\big)\big(1-\lambda\bar\lambda\big)}{1 +\lambda\bar\lambda+ 2m \big(u\bar\lambda + \bar u \lambda \big)}    \right]= \partial_t\left[ -H {\cal L}+ 4m  \frac{\big( \epsilon \psi u - \bar\epsilon\bar\psi \bar u\big)\big(1-\lambda\bar\lambda\big)}{1 +\lambda\bar\lambda+ 2m \big(u\bar\lambda + \bar u \lambda \big)}    \right].
\ee
Thus, the action \p{SusyAct1} is invariant under both the broken $S$ and unbroken $Q$ supersymmetries, and it is the
action of the $N=(2,0)$ AdS$_{3}$ superparticle.

The AdS$_3$ superparticle action \p{SusyAct1} may be written in  terms of the  Cartan forms
evaluated at $\theta=d\theta=0$ in a rather simple way as
\be\label{ActCF1}
S_0= \frac{m_0}{m} \int \Omega_J |_{\theta=0} \;.
\ee
\vspace{0.25cm}

\noindent{\bf Supersymmetric AdS$_3$ anyon}

The supersymmetrization of the anyonic-like action \p{anyonbosaction} is more involved. To construct it, let us firstly note
that, with respect to broken supersymmetry, the covariant time derivatives of the fermionic components $\cD_t \psi$ and
$\cD_t \bpsi$ transform as
\bea\label{ScDtpsitr}
&&\delta_S \cD_t \psi = 2\im m \varepsilon \, e^{2\im m t} \big( 1 -2 m \psi\bpsi   \big) -4 m \varepsilon \, e^{2\im m t} \bpsi \cD_t \psi, \nn \\
&&\delta_S \cD_t \bpsi = - 2\im m \bar\varepsilon \, e^{-2\im m t} \big( 1 -2 m \psi\bpsi   \big) +4 m \bar\varepsilon \, e^{-2\im m t} \psi \cD_t \bpsi.
\eea
Therefore, the $S$-invariant  fermionic correction to the bosonic action \p{anyonbosaction} has the form
$$
\int dt \;\cE {\cal F}\big(\lambda,\bar\lambda,u,\bar u \big) \big(  1 +4 m \psi \bpsi    \big)\big(  \cD_t \psi -2\im m \psi \big)\big( \cD_t \bpsi +2\im m  \bpsi\big).
$$
Thus,  our ansatz for the supersymmetric AdS$_3$ anyonic action reads
\be\label{sany1}
S_{anyon} = \int dt \; \cE \left[ \im \frac{\cD_t\lambda \bar\lambda - \cD_t\bar\lambda \lambda}{1-\lambda\bar\lambda} + {\cal F}\big(\lambda,\bar\lambda,u,\bar u \big) \big(  1 +4 m \psi \bpsi    \big)\big(  \cD_t \psi -2\im m \psi \big)\big( \cD_t \bpsi +2\im m  \bpsi\big)\right] \equiv \int dt \; {\cal L}_{anyon},
\ee
where the function ${\cal F}$ has to be determined by invariance under unbroken supersymmetry. The first term in \p{sany1} is a direct supersymmetrization of the bosonic anyon action and, by construction, is invariant under
broken supersymmetry.

Due to the transformation property of all our ingredients, which roughly takes the form
$$
\delta_Q \cE \sim - \partial_t \left( H \cE \right)+\ldots,\;
\delta_Q \left( u, \bu, \lambda, \blambda\right) \sim - H \left( \partial_t u, \partial_t\bu, \partial_t\lambda, \partial_t\blambda\right)+\ldots
$$
the $H$-dependent terms convert into full time derivatives. Hence, while checking invariance of the action, these terms
can be ignored.

The simplest way to fix the function ${\cal F}$ is to consider the variation of the action \p{sany1} to first order
in the fermions $\psi, \bpsi$. At this order the variation of the integrand in \p{sany1} reads (we write only the $\epsilon$-part of transformations)
\be\label{var1}
\delta_Q {\cal L}_{anyon}\approx 4 \im \epsilon \big( \partial_t \psi -2\im m \psi  \big) \partial_t \lambda \frac{1+2 m u \bar\lambda}{1+2m \bar u \lambda }\; \frac{1 + \lambda\bar\lambda +2 m \big( u\bar\lambda + \bar u \lambda   \big)}{\big(  1 -\lambda\bar\lambda \big)^2}
-\im \epsilon \big( \partial_t \psi -2\im m \psi   \big) \partial_t \lambda \cdot \frac{1 - 4 m^2 u\bar u}{\big( 1+ 2 m \bar u\lambda  \big)^2} {\cal F}.
\ee
To cancel this variation one has choose ${\cal F}$ as
\be\label{calF}
{\cal F} = 4 \frac{\big( 1 +2m \bar u\lambda   \big)\big( 1+2m u \bar\lambda   \big)\big(1 +\lambda\bar\lambda +2m \big(   u\bar\lambda + \bar u \lambda\big)    \big) }{\big( 1 -4m^2 u\bar u  \big)\big( 1-\lambda\bar\lambda  \big)^2 }.
\ee

Now, it is a matter of direct but slightly complicated calculation to check that the action \p{sany1} is invariant
under the unbroken supersymmetry to all orders in the fermionic variables. The terms which are not explicitly
cancelled in the variation of the integrand in \p{sany1} read (all terms coming from $\partial_{\lambda} {\cal F}$ and
$\partial_{\blambda} {\cal F}$ cancelled trivially)
\bea\label{Qanyontr3}
\delta_Q {\cal L} _{anyon} = -\partial_t \big( H {\cal L } _{anyon}    \big) +4m\epsilon u \psi \cE\cdot\cD_t \left[    {\cal F} \big(  1 +4 m \psi \bpsi    \big)\big(  \cD_t \psi -2\im m \psi \big)\big( \cD_t \bpsi +2\im m  \bpsi\big)\right]  \nn \\
+\epsilon \psi \cD_t\psi \cD_t \bpsi  \cE\left\{ 2\im \frac{\partial {\cal F}}{\partial \bar u}\big(  1-4m^2 u\bar u\big) +4 \im m {\cal F} \left[ - \lambda \frac{1-4m^2 u\bar u}{1+2m\bar u\lambda} - 8 \im \cD_t u + 8 m u \right]  \right\} \\
+ \epsilon \psi \bpsi \cD_t \psi \cE \cdot 4m  \left\{  \frac{\partial {\cal F}}{\partial \bar u}\big(  1-4m^2 u\bar u\big) -2 m {\cal F} \left[ \frac{\lambda+2m u}{1+2m \bar u \lambda} +2 \im \cD_t u - 4 m u    \right]  \right\}.\nn
\eea
It is straightforward to evaluate the curved brackets, as the function ${\cal F}$ is already known \p{calF}. Substituting here the expression for $\cD_t u$ \p{Qtrdu}, one finds that the last two lines in \p{Qanyontr3} combine to
\be\label{Qanyontr4}
-16 \im m \lambda \cE \frac{\big( 1 +2m \bar u\lambda   \big)\big( 1+2m u \bar\lambda   \big)}{\big( 1-\lambda\bar\lambda  \big)^2} \left[ \epsilon\psi  \cD_t\psi \cD_t \bpsi -2\im m \epsilon \psi\bpsi \cD_t \psi  \right].
\ee
These terms identically cancel the last term in the first line in \p{Qanyontr3} (after integrating by parts and substituting expression for $\cD_t u$). Thus, the anyonic action \p{sany1}, with ${\cal F}$ given by \p{calF}, is invariant with respect to both supersymmetries.

Similarly to the action of the AdS$_3$ superparticle \p{ActCF1}, the anyonic action has a rather simple shape when written
in terms of Cartan forms:
\be\label{ActCF2}
S_{anyon} =- \int \left( \Omega_3 - 4 \frac{\Omega_S \bOmega_S}{\Omega_P}\right)|_{\theta=0}.
\ee

The supersymmetric extension of the rigid particle action \p{rigid} is more complicated task. The corresponding action
can not be written in the terms of Cartan forms. The exact form of the action will be considered elsewhere.

\setcounter{equation}{0}
\section{Reductions to the nonrelativistic case}
The actions we constructed have quite a complicated structure. A possible way to simplify  our system is considering  the nonrelativistic limit. Before reducing the full supersymmetric case,
let us shortly discuss  possible reductions of the purely bosonic system, i.e. the nonrelativistic reductions of the AdS$_3$ algebra.
\subsection{Bosonic reductions}
In the bosonic case we have three possible nonrelativistic reductions of the AdS$_3$ algebra \p{bosAdS}:
\begin{itemize}
\item The first reduction consists in the following rescaling of the generators,
\be\label{rescale1}
 Z \rightarrow \omega Z, \; \bZ \rightarrow \omega \bZ, \; T \rightarrow \omega T, \; \bT \rightarrow \omega \bT.
\ee
and then taking the limit $\omega \rightarrow \infty$.  After performing of this step, we will finish with the following
algebra
\bea\label{NHooke}
&& \big[ J_3 , T  \big] = T, \; \big[ J_3 , \bT    \big] = -\bT, \quad \big[ J_3, Z \big] = Z, \; \big[ J_3, \bZ \big] = - \bZ, \nn \\
&& \big[ P, Z \big] =2 m Z, \quad \big[ P, \bZ \big] = -2 m\bZ, \quad \big[ T, P \big] = -Z , \quad \big[ \bT, P\big] = \bZ .
\eea
Clearly, the generator $J$ is decoupled from the algebra while $J_3$ generates outer automorphisms.

The algebra \p{NHooke} is just the Newton-Hooke algebra \cite{NH1,NH2}. To bring the commutation relations to the conventional
form \cite{GP, ABGR, BGKPRZ} one has to redefine the generators as follows,
\be\label{def1}
H=P-m J_3,\quad  p=Z-m T,\; {\bar p}=\bZ- m \bT,\quad G=T,\; \bG=\bT, \quad J_3.
\ee
In terms of these generators the non-zero commutators acquire a standard form,
\bea\label{NHred1}
&& \big[ H , G  \big] = p, \; \big[ H  , \bG \big] = -\bar{p}, \quad \big[ H, p  \big] = m^2 G,\;
\big[ H, \bar{p}  \big] = - m^2 \bG, \nn \\
&& \big[ J_3, p \big] = p,\; \quad \big[ J_3, \bar{p} \big] = - \bar{p},\quad
\big[ J_3, G \big] = G, \; \big[ J_3, \bG \big] = - \bG.
\eea
\item The second reduction includes an additional rescaling of the generator $J$,
\be\label{rescale2}
Z \rightarrow \omega Z, \; \bZ \rightarrow \omega \bZ, \; T \rightarrow \omega T, \; \bT \rightarrow \omega \bT, \; J \rightarrow \omega^2 J.
\ee
Taking now the limit $\omega \rightarrow \infty$ we  finish with the same relations \p{NHooke} and three new non-zero
commutators,
\be\label{NHooke2}
\big[ Z , \bZ  \big] = 4 m^2 J, \quad \big[ T , \bZ \big] = 2 m J, \; \big[ \bT, Z  \big] =- 2 m J.
\ee
The generator $J$ now becomes the central charge generator, and the corresponding algebra is
just the Bargmann-Newton-Hooke algebra \cite{GP}, i.e. the central charge extension of the Newton-Hooke algebra. In terms of the generators \p{def1} new non-zero commutators have the form
\be\label{NHred2}
\big[ p , \bG  \big] = 2 m J, \quad \big[ \bar{p} , G \big] = -2 m J.
\ee
\item The third reduction includes the rescaling of the both generators $J$ and $J_3$,
\be\label{rescale3}
Z \rightarrow \omega Z, \; \bZ \rightarrow \omega \bZ, \; T \rightarrow \omega T, \; \bT \rightarrow \omega \bT, \; J \rightarrow \omega^2 J, \; J_3 \rightarrow \omega^2 J_3.
\ee
Again, after taking the limit $\omega \rightarrow \infty$, we obtain the following commutation relations,
\bea\label{newNHooke}
&&  \big[ T, P \big] = -Z , \quad \big[ \bT, P\big] = \bZ, \quad \big[ P, Z \big] =2 m Z, \quad \big[ P, \bZ \big] = -2 m\bZ,   \nn \\
&& \big[ T, \bT  \big] =-2J_3,\quad \big[ T,\bZ\big] = 2 m J, \quad \big[ \bT, Z \big] =-2m J, \quad \big[ Z,\bZ\big] = 4 m^2 J.
\eea
Now both generators $J$ and $J_3$ become  central elements in the algebra \p{newNHooke}. Moreover, the $U(1)$ rotations
generated previously by the generator $J_3$ disappeared from the algebra. Clearly, the corresponding generator $V_3$ with
the relations
\be\label{V3}
\big[ V_3 , T  \big] = T, \; \big[ V_3 , \bT    \big] = -\bT, \quad \big[ V_3, Z \big] = Z, \; \big[ V_3, \bZ \big] = - \bZ,
\ee
can be easily added to the algebra \p{newNHooke}.

To pass to a conventional form of the algebra one has to redefine the generators as follows,
\be\label{def3}
H=P-m V_3,\quad  p=Z-m T,\; {\bar p}=\bZ- m \bT,\quad G=T,\; \bG=\bT, \quad J, \; J_3.
\ee
The full  set of non-zero commutators read
\bea\label{NHred3}
&& \big[ H , G  \big] = p, \; \big[ H  , \bG \big] = -\bar{p}, \quad \big[ H, p  \big] = m^2 G,\;
\big[ H, \bar{p}  \big] = - m^2 \bG, \nn \\
&& \big[ V_3, p \big] = p,\; \quad \big[ V_3, \bar{p} \big] = - \bar{p},\quad
\big[ V_3, G \big] = G, \; \big[ V_3, \bG \big] = - \bG, \nn \\
&& \big[ p , \bar{p}  \big] = -2 m^2 J_3,\; \big[ p , \bG  \big] = 2m \left( J+J_3\right),\;
\big[ \bar{p} , G  \big] = -2m \left( J+J_3\right),\;
\big[ G , \bG  \big] = -2 J_3,
\eea
Thus, we have two central charge extensions of the Newton-Hooke algebra \cite{gen1}. The extension with the
central charge $J_3$ exists in three-dimensional space-time only and it called ``exotic'' central extension \cite{exotic}.
\end{itemize}
It should be clear that if we choose the coset element in the usual way as
\be\label{newNHooke-coset}
g = e^{\im t P} e^{\im (uZ + \bar u \bZ)} e^{\im (\lambda T + \bar\lambda \bT)}
\ee
then the central charges in the algebra \p{NHooke2} and in the algebra \p{newNHooke} will not have any realization on the
coordinates and fields, while the $Z, \bZ$ and $T, \bT$ transformations will be realized in the same way for all three algebras:
\be
\delta_Z u = e^{-2 \im m t} \, a, \quad \delta_Z \bu = e^{2 \im m t} \, \bar{a}, \qquad
\delta_T u =\frac{e^{-2 \im m t} -1}{2m} \, b, \quad \delta_T \bu =\frac{e^{2 \im m t} -1}{2m} \, \bar{b}. \label{ZTtr}
\ee
The advantage of the third reduction is the presence of two new Cartan forms for the central charge generators which,
as we will see shortly, can be used to construct invariant actions.

With the coset element \p{newNHooke-coset} the Cartan forms read
\bea\label{newNHooke-cartan}
&& \omega_P = dt, \quad  \omega_T = d\lambda ,  \quad \bar\omega_T = d\bar\lambda, \quad \omega_3 = \im \big( \lambda\, d\bar\lambda -\bar\lambda\, d \lambda  \big),\nn \\
&&\omega_Z = du + 2\im m u\,dt + \im \lambda\, dt , \quad \bar\omega_Z = d\bar u - 2\im m \bar u\, dt - \im \bar\lambda\, dt ,    \nn \\
&& \omega_J = - 8 m^3 u\bar u dt + 2\im m^2 (\bar u d u - d \bar u u) -2 m \lambda \bar\lambda dt - 2\im m \left[ \lambda (d\bar u - 2\im m \bar u) - \bar\lambda (d u + 2\im m u)    \right].
\eea
Similarly to the previously considered bosonic case, one may impose the inverse Higgs effect conditions \p{ih}
$$
\omega_Z = \bar\omega_Z =0
$$
which result in  expressing $\lambda, \blambda$ in terms of $u$ and $\bu$,
\be\label{lambdared}
\lambda = \im \dot{u} - 2 m u, \quad \blambda = - \im \dot{\bu}  - 2 m \bu.
\ee
Keeping in the mind that the action $\int \omega_P = \int dt$ is trivial, we have  three possible invariant actions:
\bea\label{redact}
&& S_0^{bos} = \frac{1}{2m} \int \omega_J =\int dt \left[ \dot{u} \dot{\bu} - \im m \left( \dot{u} \bu - u \dot{\bu} \right) \right], \\
&& S_{anyon}^{bos} =- \int \omega_3 =  \int dt \left\{ \im \left[ \ddot{u}\dot{\bu} -\dot{u}\ddot{\bu} + 4 m^2\left( \dot{u} \bu -u \dot{\bu}\right)\right] - 8 m \dot{u} \dot{\bu} \right\},\\
&& S_{rigid}^{bos} = \int \frac{\omega_T {\bar\omega}_T}{\omega_P} =  \int dt \left[ \ddot{u} \ddot{\bu} - 2 \im m \left( \ddot{u}\dot{\bu} -\dot{u} \ddot{\bu}\right)+
4 m^2 \dot{u} \dot{\bu} \right].
\eea

These actions may be slightly simplified by passing to the new variables
\be\label{qqb}
q = e^{-\im \gamma m t} u, \quad \bq =e^{\im \gamma m t} \bu,
\ee
 in which they acquire the form
\bea\label{qactions}
S_0^{bos} &=& \int dt \left\{ \dot q \dot{\bar q} - \im m (1+\gamma) \big( \dot q \bar q - \dot{\bar q} q   \big) + m^2 \gamma \big(  2 + \gamma\big) q\bar q   \right\}, \nn \\
S_{anyon}^{bos} &=& \int dt \left\{\im \big( \ddot q \dot{\bar q} - \ddot{\bar q} \dot q \big) - 2 m (4+3 \gamma) \dot q \dot{\bar q} - 2 m^3 \gamma ( 2+\gamma  )^2 q\bar q  \right. \nn \\&& \left. +\im  m^2 \big( 4+8\gamma + 3 \gamma^2    \big) \big( \dot q \bar q - \dot{\bar q} q \big)  \right\},\nn\\
S_{rigid}^{bos} &=&  \int dt \left\{ \ddot q \ddot{\bar q} -2\im m (1+\gamma)\big( \ddot q \dot{\bar q} - \ddot{\bar q} \dot q  \big) + 2 m^2 \big(2+6 \gamma+3 \gamma^2\big) \dot q \dot{\bar q} \right. \nn \\ &&\left.  -2\im \gamma ( 1+\gamma  ) (2+\gamma)m^3 \big( \dot q \bar q - \dot{\bar q} q \big) + m^4 \gamma^2
\big(2 + \gamma\big)^2 q\bar q  \right\}.\nn
\eea
Choosing, for example, $\gamma = -1$ one may bring the action $S_0^{bos}$ to the  harmonic oscillator \cite{BGKPRZ},
\be\label{BelNer}
S_0^{bos} =  \int dt \left( \dot q \dot{\bar q} - m^2  q\bar q   \right),
\ee
as it should be for a Newton-Hooke particle.
Then the actions $S_{anyon}^{bos}$ and  $S_{rigid}^{bos}$ with $\gamma = -1$ will give corresponding higher derivative corrections to this action. All these actions are invariant under the algebra \p{newNHooke}.

\subsection{Supersymmetric nonrelativistic Newton-Hooke particle}
It is possible to construct a nonrelativistic version of the $N=(2,0)$ supersymmetric AdS$_3$ algebra and the corresponding superparticle actions.
Remembering our rescaling of the bosonic subalgebra \p{rescale2} and preserving the unbroken supersymmetry, one can  not rescale the generators $Q,\bQ$. In addition, for keeping the relation $\big\{ Q, S  \big\} = 2\bZ$ in order to realize some sort of inverse Higgs effect, it is required to rescale the $S$-generator as $S \rightarrow \omega S$. Performing such a rescaling and taking the limit $\omega \rightarrow \infty$, we  obtain the following superalgebra,
\bea\label{SUSYNHooke}
&& \big\{ Q,\bQ \big\} = 2 P, \quad, \big\{Q, S\} =2 \bZ, \quad \big\{ \bQ , \bS  \big\} = 2Z, \quad \big\{ S,\bS  \big\} =-4m J, \nn \\
&& \big[ Z, Q    \big] = -2 m\bS, \quad \big[ \bZ, \bQ \big] = 2m S, \quad \big[ P, S \big] = -2 mS, \quad \big[ P, \bS \big] = 2m \bS,   \\
&& \big[ T, Q   \big] = -\bS, \quad \big[ \bT, \bQ   \big] = S, \quad \big[ T, P \big] = -Z , \quad \big[ \bT, P\big] = \bZ, \quad \big[ P, Z \big] =2 m Z, \quad \big[ P, \bZ \big] = -2 m\bZ,  \nn \\
&& \big[ T, \bT  \big] =-2J_3,\quad \big[ T,\bZ\big] = -2 m J, \quad \big[ \bT, Z \big] =2m J, \quad \big[ Z,\bZ\big] = 4 m^2 J,\nn
\eea
where the generators $J$ and $J_3$ are still central elements of the superalgebra.

The coset element can be parameterized as before \p{coset},
$$
g = e^{\im t P} e^{\theta Q + \bar\theta \bQ} e^{\mpsi S + \mbpsi \bS} e^{\im (\mmu Z + \mmbu \bZ)} e^{\im (\mlambda T + \mblambda \bT)}.
$$
With such a coset element the Cartan forms are simplified to
\bea\label{SUSYNHooke-forms2}
\omega_P & = & \triangle t = dt - \im \big( \theta d\bar\theta + \bar\theta d \theta \big), \quad \omega_Q = d\theta, \quad
\omega_T = d\mlambda,  \;
 \omega_3 = \im \big( \mlambda d\mblambda - \mblambda d \mlambda  \big),  \nn \\
\omega_Z & = & d \mmu  + \im \big(2m \mmu + \mlambda \big) \triangle t -2\im \mbpsi d\bar\theta, \quad
\omega_S  =  d\mpsi -2 \im m \mpsi \triangle t     -\im \big(  2 m \mmbu + \mblambda \big)d\bar\theta,  \nn \\
\omega_J &=& 2\im m \big( \mbpsi d \mpsi + \mpsi d\mbpsi    \big) - 8 m^2 \mpsi\mbpsi \triangle t - 8 m^2 \big( \mmu \mpsi d\theta - \mmbu \mbpsi d\bar\theta \big) - 8 m^3 \mmu \mmbu \triangle t  -2 m \mlambda \mblambda \triangle t \nn \\
&& -2\im m \mlambda \big(d \mmbu  - 2 \im m \mmbu \triangle t -2\im \mpsi d\theta      \big) +2 \im m \mblambda \big(  d \mmu  + 2 \im m \mmu \triangle t -2\im \mbpsi d\bar\theta  \big)  -2\im m^2 \big( \mmu d\mmbu - \mmbu d\mmu    \big).
\eea
Imposing the standard conditions of the inverse Higgs effect \p{ih},
$$\omega_Z =0, \, \bar\omega_Z = 0,$$
we get the equations
\bea
&& D \mmu =0, \quad \bD \mmbu =0, \label{chir} \\
&& \mpsi = \frac{\im}{2} D\mmbu,\; \mbpsi = \frac{\im}{2} \bD\mmu, \quad
\mlambda = \im \big( \dot \mmu + 2\im m \mmu   \big), \; \mblambda = -\im \big( \mathbf{\dot{\bar{u}}} - 2\im m \mmbu   \big),
\eea
where the flat spinor derivatives $D, \bD$ were defined in \p{cD1}. We are dealing with  chiral superfields $\mmu, \mmbu$ and, therefore,
the physical component fields may be defined in the usual way as
\be
u = \mmu |_{\theta\rightarrow 0}, \, \bar u = \mmbu |_{\theta\rightarrow 0},\quad
\psi = \mpsi |_{\theta\rightarrow 0}, \, \bpsi = \mbpsi |_{\theta\rightarrow 0}.
\ee
The transformation properties of these components under the unbroken supersymmetry $Q$ and the broken supersymmetry $S$ are very simple,
\bea
&& \delta_Q u = -2\im \bar\epsilon \bpsi, \; \delta_Q \bar u = -2\im \epsilon \psi, \quad \delta_Q \psi = \bar\epsilon \dot{\bar u}, \; \delta_Q \bpsi = \epsilon \dot u, \label{Qtr} \\
&&\delta_S u = 0, \; \delta_S \bu = 0, \quad
\delta_S \psi = \varepsilon e^{2\im m t}, \; \delta_S \bpsi = \bar\varepsilon e^{-2\im m t} . \label{Str}
\eea

To construct the invariant supersymmetric actions generalizing the bosonic actions \p{action1}, \p{anyonu}, \p{rigid}, we must investigate $S$ and $Q$ supersymmetries.
It is not hard to check that these actions are $S$-invariant but not $Q$-invariant.  Thus, one has to find fermionic completions invariant
under $Q$ supersymmetry. It is rather easy to see that the following fermionic actions,
\bea\label{SUSYredact}
&& S_0^{ferm} =  \int dt \left[ \im \big( \dot\psi \bpsi + \dot\bpsi \psi \big)+ 4m \psi\bpsi \right], \\
&& S_{anyon}^{ferm}  = \int dt \left[ \dot\psi\dot\bpsi - 4 m^2 \psi\bpsi \right],\\
&& S_{rigid}^{ferm} = \int dt \left[ \im \big( \ddot\psi \dot\bpsi + \ddot\bpsi \dot\psi \big) + 16 m^3 \psi\bpsi\right]
\eea
have the proper dimension and are $S$-invariant. Thus, our ansatz for the fully supersymmetric actions is
\bea\label{SUSYredactA}
&& S_0 = S_0^{bos}+\gamma_0 S_0^{ferm}, \\
&& S_{anyon} = S_{anyon}^{bos} +\gamma_1 S_{anyon}^{ferm} + m \gamma_2 S_0^{ferm}, \\
&& S_{rigid} = S_{rigid}^{bos} + \gamma_3 S_{rigid}^{ferm}+ m \gamma_4 S_{anyon}^{ferm} + m^2 \gamma_5 S_0^{ferm},
\eea
where $\gamma_0, \ldots,\gamma_5$ are  constant parameters. Imposing  invariance of these actions under the transformations \p{Qtr},
these constants can be uniquely fixed as
\be\label{sol1}
\gamma_0 = -1,\quad \gamma_1 =4,\; \gamma_2=8,\quad \gamma_3 =-1,\; \gamma_4=-8,\;\gamma_5=-4.
\ee
Thus, the full actions invariant under both the broken $(S)$ and the unbroken $(Q)$ supersymmetries  read
\bea\label{finactionsR}
S_0 &=& \int dt \big\{  \dot u \dot{\bar u} -  \im m \big( \dot u \bar u - \dot{\bar u} u  \big) - \im \big( \dot \psi \bpsi- \psi \dot\bpsi   \big) - 4 m \psi\bpsi  \big\}, \nn \\
S_{anyon} &=&\int dt \left[\im \left( \ddot{u}\dot{\bu} -\dot{u}\ddot{\bu}\right)- 8 m \dot{u} \dot{\bu} + 4\im m^2\left( \dot{u} \bu -u \dot{\bu}\right)  +4 \dot\psi \dot{\bpsi} + 8 \im m \left( \dot\psi \bpsi -\psi\dot{\bpsi}\right) + 16 m^2 \psi\bpsi \right],\nn \\
S_{rigid} &=& \int dt \left[ \ddot{u} \ddot{\bu} - 2 \im m \left( \ddot{u}\dot{\bu} -\dot{u} \ddot{\bu}\right)+
4 m^2 \dot{u} \dot{\bu} - \im \left( \ddot\psi \dot\bpsi -\dot\psi \ddot\bpsi \right) -8 m \dot\psi\dot\bpsi - 4\im m^2 \left( \dot\psi \bpsi -\psi\dot\bpsi\right) \right].
\eea

Note that the Newton-Hooke superparticle actions $S_0$ and $S_{anyon}$ may be represented, similarly to the bosonic case, as integrals of the Cartan forms  \p{SUSYNHooke-forms2} taken in the $\theta \rightarrow 0$ limit:
\be\label{an}
S_0 = \frac{1}{2m} \int \; \omega_J|_{\theta \rightarrow 0}, \quad
S_{anyon} = -\int \omega_3|_{\theta \rightarrow 0}+4 \int \left( \frac{\omega_S {\bar\omega}_S}{\omega_P}\right)_{\theta \rightarrow 0} .
\ee

Finally, the bosonic part of these actions may be slightly simplified, as in \p{qactions}, by passing to the $q, \bq$ variables \p{qqb}. In addition, one
may  redefine the fermionic components in a similar way as
\be\label{ksi0}
\psi = e^{-\im \rho m t} \xi, \quad \bpsi =e^{\im \rho m t} \bar\xi.
\ee
Then fixing the parameter $\rho$ one may reach the desired form of the fermionic part of the actions. For example, choosing the new fermionic variables as
\be\label{ksi}
\xi = e^{ -2 \im m t} \psi, \quad \bxi =e^{2 \im m t} \bpsi,
\ee
one may represent the fermionic parts of the actions \p{finactionsR} as
\be\label{fermsimp}
S_0 = \int dt \left[ \ldots - \im \big( \dot \xi \bxi- \xi \dot\bxi   \big)\right], \;
S_{anyon} =\int dt \left[\ldots +4 \dot\xi \dot{\bxi}\right], \;
S_{rigid}-m S_{anyon}  =\int dt \left[\ldots - \im \left( \ddot\xi \dot\bxi -\dot\xi \ddot\bxi \right)\right].
\ee

\newpage

\subsection{Comments}
We make some comments concerning the nonrelativistic case.
\begin{itemize}
\item The structure of the bosonic actions \p{redact} is completely fixed by the symmetry under the $(Z, \bZ)$ and $(T, \bT)$
transformations realized as \p{ZTtr},
$$
\delta_Z u = e^{-2 \im m t} \, a, \quad \delta_Z \bu = e^{2 \im m t} \, \bar{a}, \qquad
\delta_T u =\frac{e^{-2 \im m t} -1}{2m} \, b, \quad \delta_T \bu =\frac{e^{2 \im m t} -1}{2m} \, \bar{b}.
$$
\item The fermionic components $(\psi, \bpsi)$ are invariant under the $(Z, \bZ)$ and $(T, \bT)$
transformations
\be\label{ZTtr0}
\delta_Z \psi = \delta_Z \bpsi =0, \qquad \delta_T \psi = \delta_T \bpsi =0
\ee
and, thus, the fermionic terms completing the bosonic actions  to the full supersymmetric ones \p{finactionsR} are determined by their
invariance under  the $S, \bS$ and $Q, \bQ$ transformations \p{Qtr}, \p{Str}.
\item The action $S_0$ \p{finactionsR} is a some variant of generalized Galileon action (see e.g. \cite{GHJT,KO}). Indeed, the transformations \p{ZTtr}
are some variant of polynomial shift symmetries \cite{GGHY}. They go to the standard Galileon symmetries in the flat $(m \rightarrow 0)$ limit
\be
\delta_Z u|_{m \rightarrow 0} = a, \qquad \delta_T u|_{m \rightarrow 0} = - \im b\,t .
\ee
\item Our supersymmetric actions \p{finactionsR} are a supersymmetrization of extended Galileon actions. Of course, these are not really Galileons,
because in one dimension the higher-dimensional terms in the action always produce  higher-order equations of motion. Nevertheless, our
actions possess the proper extension \p{ZTtr} of the Galilean symmetries \p{ZTtr0}. Funnily enough, the simplest harmonic oscillator action \p{BelNer} (as well as its supersymmetric extension) features such a symmetry, and it may be called {\it ``extended Galileon''}.
\item Due to the slightly non-standard reduction we used \p{rescale2}, the actions \p{an} can be represented in terms
of Cartan forms which behave as  Wess-Zumino terms under $(Z,T,S)$ transformations. Performing the first reduction \p{rescale1}, the
forms $\omega_J, \omega_3$ will vanish and, therefore, the proper actions should be constructed in a standard way as  Wess-Zumino terms \cite{GHJT}.
\end{itemize}

\setcounter{equation}{0}
\section{Conclusion}
We have extended our previous analysis of a superparticle moving in flat $D=2{+}1$ spacetime (including higher time derivatives) \cite{kkln} to a superparticle moving on AdS$_3$, with $N{=}(2,0)$, $D{=}3$ supersymmetry partially broken to $N{=}2$, $d{=}1$. We have employed the coset approach to constructing the component actions. The higher time-derivative terms were chosen to preserve all (super)symmetries of the free superparticle in AdS$_3$. The actions have a nice form in terms of covariant Cartan forms. We also considered the nonrelativistic limit, in which
our superalgebra turns into the Newton-Hooke superalgebra extended by two central charges, and the reduced actions describe a Newton-Hooke superparticle including higher derivative terms.

Our consideration was purely classical. To analyze the effects of the higher derivative terms one should quantize the system.
In the present case this is much more complicated than for a superparticle moving in flat $D=2{+}1$ spacetime.
Already the quantization of the nonrelativistic systems constructed here should be quite useful. We are planning to come back to this task in future publications.

\setcounter{equation}{0}
\section*{Acknowledgments}
We are grateful to Anton Galajinsky and Armen Nersessian for valuable correspondence.

\noindent The work of N.K. and S.K. was partially supported by RSCF grant 14-11-00598 and by \\
RFBR grant 15-52-05022~Arm-a.


\end{document}